\documentclass[ reprint,nofootinbib,amsmath,amssymb,onecolumn,aps]{revtex4-2}
\usepackage{appendix}
\usepackage{graphicx}
\usepackage{dcolumn}
\usepackage[colorlinks,linkcolor=magenta,anchorcolor=cyan,citecolor=blue]{hyperref}
\usepackage{physics}
\usepackage{bm}

\def\be{\begin{equation}}
\def\ee{\end{equation}}
\def\ba{\begin{eqnarray}}
\def\ea{\end{eqnarray}}

\def\dbar{{\mathchar '26\mkern -10mu\delta}}

              \def\.{\cdot}

\begin{document}
\title{Background subtraction method is not only much simpler, but also as applicable as covariant counterterm method}
\author{Wei Guo$^{1,2}$}
\email{guow@mail.bnu.edu.cn}
\author{Xiyao Guo$^{1,2}$}
\email{xiyaoguo@mail.bnu.edu.cn}
\author{Xin Lan$^{1,2}$}
\email{xinlan@mail.bnu.edu.cn}
\author{Hongbao Zhang$^{1,2}$}
\email{hongbaozhang@bnu.edu.cn}
\author{Wei Zhang$^{1,2}$}
\email{w.zhang@mail.bnu.edu.cn}
\affiliation{$^1$School of Physics and Astronomy, Beijing Normal University, Beijing 100875, China\label{addr1}\\
$^2$ Key Laboratory of Multiscale Spin Physics, Ministry of Education, Beijing Normal University, Beijing 100875, China
}

\date{\today}

\begin{abstract}
As the criterion for the applicability of the background subtraction method,  not only the finiteness condition of the resulting Hamiltonian but also the condition for the validity of the first law of black hole thermodynamics can be reduced to the form amenable to much simpler analysis at infinity by using the covariant phase space formalism. With this, we further establish that the background subtraction method is as applicable as the covariant counterterm method not only to Einstein's gravity, but also to its higher derivative corrections for black hole thermodynamics in both asymptotically flat and AdS spacetimes.
In addition, our framework also provides us with the first derivation of the universal expression of the Gibbs free energy in terms of the Euclidean on-shell action beyond Einstein's gravity. Among others, our findings have a far reaching impact on the shift in methodology for the Euclidean approach to black hole thermodynamics, where the well justified background subtraction method by our wieldy criterion is supposed to be the favored choice compared to the covariant counterterm method. 


\end{abstract}
\maketitle
\section{Introduction}
Since its advent, not only has the Euclidean approach to quantum gravity 
demonstrated its remarkable power in disclosing the thermal nature of stationary black holes, but also displayed its great success in calculating the corresponding Gibbs free energy and related thermodynamic quantities by either the background subtraction method \cite{Hartle,GH,HP,HH,GPP} or the covariant counterterm method \cite{BFS,PS,MM}. In particular, via the Euclidean quantum gravity, or its variant AdS/CFT correspondence for asympotically AdS spacetimes, significant progress has recently been made in our understanding of the higher derivative corrections to black hole thermodynamics \cite{RS,Cheung,Cremonini,Belgium,Melo,Bobev,Xiao1,Cassani,Noumi,Ma,Zatti,Hu,Ma2,Massai,Xiao2}, which not only supplies some universal information about the unknown UV theory within the framework of effective field theory, but also sheds insight into the dynamics of the boundary quantum systems beyond the strong coupling and infinite $N$ limits in the context of AdS/CFT correspondence. Despite this, 
the only two available methods mentioned above are both suspected to have their disadvantages, respectively. The covariant counterterm method turns out to be unwieldy compared to the background subtraction method, because the former generically needs additional terms added on the boundary on top of the generalized Gibbons-Hawking term for a general diffeomorphism covariant theory of gravity. It is noteworthy that these additional boundary terms are not only highly complicated, but also dependent delicately on the spacetime dimension. While the background subtraction method is believed to have a rather restrictive applicability in the sense that the induced metric of the stationary black holes on the boundary cannot always be embedded isometrically into the reference spacetime \cite{PS,MM,Ma,Hu,Ma2}. But nevertheless, as evidenced miraculously by the final correct result it gives rise to,  the background subtraction method seems also applicable to some specific circumstances where such an embedding is impossible. This puzzle has remained elusive since the background subtraction method was proposed long before back in the 1970s. 

Among others, one main purpose of this paper is to resolve the above longstanding puzzle in a confirmative way. To this end, 
for the first time we have succeeded in reducing the necessary and sufficient conditions for the applicability of the background subtraction method into the form amenable to much simpler analysis at infinity by using the covariant phase space formalism \cite{Wald,IW,Iyer,
DG,Liberati,Harlow,Zhang,Zhang1,GSTV} in an innovative manner. It turns out that the resulting criterion does not require the induced metrics be the same, thus evades the restriction on the applicability of the background subtraction method from the isometric embedding. Based on our newly formulated criterion, we establish in an explicit way that the background subtraction method is actually as applicable as the covariant counterterm method not only to Einstein's gravity but also to its higher derivative corrections in both asymptotically flat and AdS spacetimes, overturning the prevailing belief that the former is inferior to the latter in this regard. Therefore our finding offers the first universal proof for the applicability of the background subtraction method in the aforementioned more generic circumstances. 
In passing, we also fill the gap in all the calculations via the background subtraction method for the higher derivative corrections to black hole thermodynamics by providing the first proof of the universal relation between the Gibbs free energy and the on-shell Euclidean action beyond Einstein's gravity, which has been previously assumed without derivation \cite{RS,Cheung,Melo,Bobev,Xiao1,Cassani,Noumi,Massai,Xiao2}. 
In particular, our work will have a direct impact on the shift in methodology for the Euclidean approach to black hole thermodynamics, where the well justified background subtraction method by our wieldy criterion is supposed to be the favorable choice for the relevant calculations, which otherwise would become much more involved via the covariant counterterm method. 




We will follow the notation and conventions of \cite{GR}. In particular, we shall use the boldface letters to denote differential forms with the tensor indices suppressed. In addition, we shall work in the units with $c=\hbar=k_B=G_N=1$.

\section{Covariant phase space formalism for $F(R_{abcd})$ gravity}
In this section, we shall present the relevant ingredients of covariant phase space formalism for $F(R_{abcd})$ gravity. For details, please refer to \cite{Zhang1}. As such, we start from the Lagrangian form for $F(R_{abcd})$ gravity as follows
\begin{equation}
\mathbf{L}=\bm{\epsilon} F(R_{abcd},g_{ab}),
\end{equation}
where $\bm{\epsilon}$ is the spacetime volume form, and $F$ is an arbitrary function of $R_{abcd}$ and $g_{ab}$. The variation of the Lagrangian form is given by 
\begin{equation}\label{variationradius}
    \delta \mathbf{L}= \bm\epsilon E^{ab}\delta g_{ab}+d\mathbf{\Theta}.
\end{equation}
Here $E^{ab}=0$ is the equation of motion with 
\begin{eqnarray}
E_g^{ab}=\frac{1}{2}g^{ab}F+\frac{1}{2}\frac{\partial F}{\partial g_{ab}}+2\nabla_{c}\nabla_{d}\psi^{c(ab)d},
\end{eqnarray}
and $\mathbf{\Theta}=\theta\cdot\bm{\epsilon}$ is called the bulk symplectic potential with 
 \begin{equation}
\theta^a=2(\nabla_d\psi^{bdca}\delta g_{bc}-\psi^{bdca}\nabla_d\delta g_{bc}),
\end{equation}
where $\psi^{abcd}$ is defined as the derivative of $F$ with respect to the Riemann tensor $R_{abcd}$ by assuming it is independent of the metric, abbreviated as
$\psi^{abcd}\equiv \frac{\partial F}{\partial R_{abcd}}$.

The Noether current associated with a vector field $\xi$ is given by 
\begin{equation}\label{Noethercurrent}
\mathbf{J}_\xi\equiv X_\xi\cdot \mathbf{\Theta}-\xi\cdot \mathbf{L}=d\mathbf{Q}_\xi,
\end{equation}
when evaluated on the on-shell solutions with 
the corresponding Noether charge
\begin{equation}\label{waldcharge}
 \mathbf{Q}_\xi=(-\psi^{cadb}\nabla_{[d}\xi_{b]}+2\nabla_{[d}\psi^{cadb}\xi_{b]})\bm{\epsilon}_{ca\cdot\cdot\cdot}.
\end{equation}
Here $X_\xi\equiv\int d^dx\sqrt{-g}\mathcal{L}_\xi g_{ab}(x)\frac{\delta }{\delta g_{ab}(x)}$ is understood as a vector at the point $g_{ab}(x)$ in the configuration space and $\delta$ is regarded as the exterior derivative in the configuration space.

On the other hand, with $n_a$ as the outward-pointing normal vector and $h_{ab}$ as the induced metric of a timelike boundary $\Gamma$, which is assumed to be fixed under variation, the variation of $n_a$ is always proportional to $n_a$, namely $\delta n_a=\delta a n_a$ with $\delta a$ the proportional coefficient. Accordingly, one can further have 
\begin{eqnarray}
    \delta n^a=-\delta a n^a-\dbar{A}^a, \quad
    \delta g^{ab}=-2 \delta a n^an^b-\dbar A^a n^b-n^a\dbar A^b+\delta h^{ab}
\end{eqnarray}
with $\dbar{A}^an_a=0$. Whence the bulk symplectic potential can be cast into the following form
\begin{eqnarray}
    \mathbf{\Theta}|_{\Gamma}=-\delta \mathbf{B}+d\mathbf{C}+\mathbf{F},
\end{eqnarray}
where
\begin{eqnarray}
    \mathbf{B}=4\Psi_{ab}K^{ab}\hat{\bm\epsilon},
    \quad \mathbf{C}=\mathbf\omega\cdot\hat{\bm\epsilon},\quad 
    \mathbf{F}=\hat{\bm\epsilon}(T_{hbc}\delta h^{bc}+T_{\Psi bc}\delta\Psi^{bc})
\end{eqnarray}
with $K_{ab}$ the extrinsic curvature, $\hat{\bm\epsilon}$ the induced volume defined as  $\bm\epsilon=\mathbf n\wedge \hat{\bm\epsilon}$, $\Psi_{ab}=\psi_{acbd}n^cn^d$, and
\begin{eqnarray}\label{bc}
   && \omega^a=-2\Psi^a{}_b\dbar A^b+2h^{ae}\psi_{ecdb}n^d\delta h^{bc},  
    \nonumber\\
    &&T_{\text{h}bc}=-2\Psi_{de}K^{de}h_{bc}+2n^a\nabla^e\psi_{deaf}h^d{}_{(b}h^f{}_{c)}-2 \Psi_{a(b}K^a{}_{c)}-2D^a(h_a{} ^eh_{(c}{}^f\psi_{|efd|b)}n^d), \quad  T_{\Psi bc}=4 K_{bc}.
\end{eqnarray}
Next let us introduce the symplectic form as follows\cite{Harlow,Zhang}
\begin{equation}
    \Omega\equiv\int_{\Sigma}\delta \mathbf{\Theta}-\int_{\partial\Sigma_1}\delta \mathbf{C}.
\end{equation}
Here we assume that the spacelike hypersurface $\Sigma$ stretches out and terminates at $\Gamma$ with the outer boundary $\partial\Sigma_1=\partial\Sigma\cap\Gamma$. In addition, the volume element on $\Sigma$ is defined as $\bm\epsilon=\mathbf{\nu}\wedge \bar{\bm\epsilon}$ with $\nu_a$ the future-directed normal vector and the volume element on $\partial\Sigma_1$ is defined as $\bar{\bm{\epsilon}}=\mathbf{n}\wedge\tilde{\bm\epsilon}$.
Then for $\xi$ tangential to $\Gamma$, one can show that
\begin{eqnarray}\label{fe1}
    -X_\xi\cdot \Omega&=&-\int_\Sigma X_\xi\cdot\delta \mathbf{\Theta}+\int_{\partial\Sigma_1}X_\xi\cdot \delta \mathbf{C}=-\int_\Sigma \mathcal{L}_{X_\xi}\mathbf{\Theta}-\delta(X_\xi\cdot\mathbf{\Theta})+\int_{\partial\Sigma_1}\mathcal{L}_{X_\xi}\mathbf{C}-\delta(X_\xi\cdot\mathbf{C})\nonumber\\
    &=&\int_\Sigma \delta(X_\xi\cdot\mathbf{\Theta})-\mathcal{L}_\xi\mathbf{\Theta}+\int_{\partial\Sigma_1}\mathcal{L}_\xi\mathbf{C}-\delta(X_\xi\cdot\mathbf{C})=\int_\Sigma\delta \mathbf{J}_\xi-d(\xi\cdot\mathbf{\Theta})+\int_{\partial\Sigma_1}\xi\cdot d\mathbf{C}-\delta(X_\xi\cdot\mathbf{C})\nonumber\\
    &=&\int_\Sigma d(\delta \mathbf{Q}_\xi-\xi\cdot\mathbf{\Theta})+\int_{\partial\Sigma_1}\xi\cdot d\mathbf{C}-\delta(X_\xi\cdot \mathbf{C})=-\int_{\partial\Sigma_0}\delta\mathbf{Q}_\xi-\xi\cdot\mathbf{\Theta}+\int_{\partial\Sigma_1}\delta(\mathbf{Q}_\xi+\xi\cdot\mathbf{B}-X_\xi\cdot\mathbf{C})-\xi\cdot\mathbf{F},\nonumber\\
\end{eqnarray}
where $\partial\Sigma_0$ denotes the inner boundary of $\Sigma$ with the same orientation as $\partial\Sigma_1$ if it exists, and the equation of motion is used in the third line.


\section{Criterion for the applicability of background subtraction method}\label{bsm}
\begin{figure}
\centering
\includegraphics[width=0.5\textwidth]{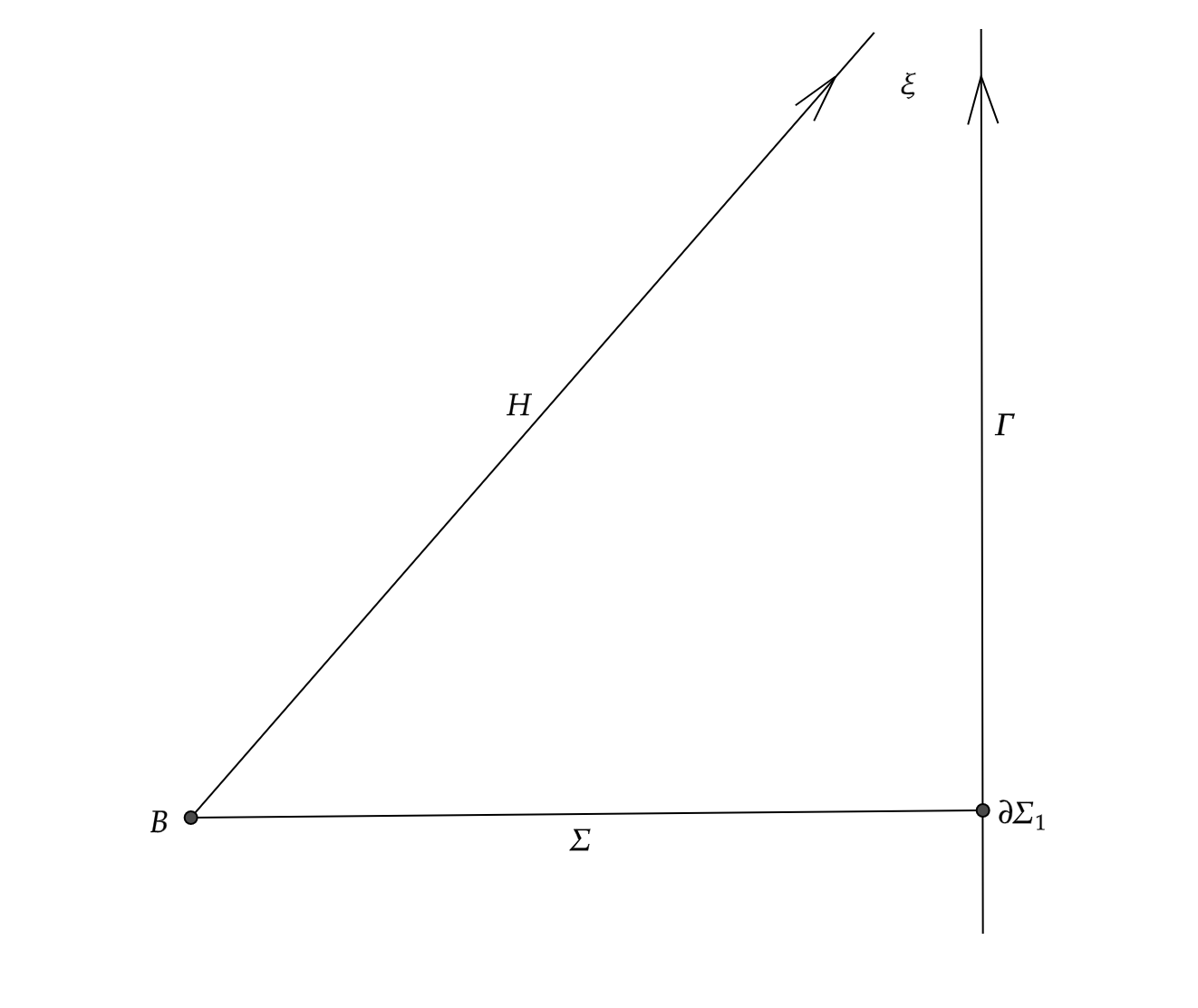}
\caption{The spacelike hypersurface $\Sigma$ emanates from the bifurcation surface $B$ of the event horizon $H$ and terminates at the timelike hypersurface $\Gamma$ with the Killing field $\xi$ tangential to both $H$ and $\Gamma$.}\label{fig1}
\end{figure}

As demonstrated in FIG. \ref{fig1}, now let us fix $\xi$ as the Killing vector field $\frac{\partial}{\partial t}$ normal to the event horizon of the stationary black hole under consideration and choose the spacelike hypersurface $\Sigma$ such that its inner boundary is simply the bifurcation surface $B$ at which $\xi$ vanishes, and its outer boundary lies on $\Gamma$ to which $\xi$ is tangential. Then Eq. (\ref{fe1}) gives rise to 
\begin{equation}\label{fls1}
    \frac{\kappa}{2\pi}\delta S=\delta H_\xi-\int_{\partial\Sigma_1}\xi\cdot \mathbf{F},
\end{equation}
where the black hole entropy is defined as 
\begin{equation}\label{waldformula}
    S\equiv\frac{2\pi}{\kappa}\int_B\mathbf{Q}_\xi,
\end{equation}
and the Hamiltonian conjugate to $\xi$ is defined as 
\begin{equation}\label{Gibbs3}
H_\xi\equiv\int_{\partial\Sigma_1}\mathbf{Q}_\xi+\xi\cdot\mathbf{B}-X_\xi\cdot\mathbf{C}.
\end{equation}
As shown in \cite{Zhang1}, the above Hamiltonian can be expressed in terms of the generalized Brown-York boundary energy momentum tensor $\mathcal{T}^a{}_c\equiv -(2T_\text{h}{}^a{}_c+2\Psi^{ab}T_{\Psi cb})$, i.e.,
\begin{equation}\label{simple}
H_\xi=\int_{\partial\Sigma_1}q_\xi\cdot\hat{\bm\epsilon}
\end{equation}
with
\begin{eqnarray}\label{brownyork}
    q_\xi^a=\mathcal{T}^a{}_c\xi^c,
    \end{eqnarray}
if $\xi$ is tangential to $\Gamma$.
Next let us move onto the background subtraction procedure. As such, we are required to choose a reference spacetime without the event horizon. For the stationary black holes in asymptotically flat spacetimes, the standard choice for the reference spacetime is the Minkowski spacetime. For the stationary black holes in asymptotically AdS spacetimes, the pure AdS spacetime serves as the standard reference spacetime. Then Eq. (\ref{fe1}) gives rise to 
\begin{equation}\label{fls0}
    0=\delta H^0_\xi-\int _{\partial\Sigma_1}\xi\cdot \mathbf{F}^0,
\end{equation}
    where the superscript $0$ denotes the corresponding quantities evaluated on the aforementioned reference spacetime. Combining Eq. (\ref{fls1}) and Eq. (\ref{fls0}), we have the first law of black hole thermodynamics 
    \begin{equation}
        \frac{\kappa}{2\pi}\delta S=\delta [H_\xi]
    \end{equation}
    with the square bracket representing the difference from the reference spacetime as $[H_\xi]=H_\xi-H^0_\xi$,
    if and only if 
    \begin{equation}\label{ourcriterion}
        \int_{\partial\Sigma_1}\xi\cdot[\mathbf{F}]=0
    \end{equation}
    as $\Gamma$ approaches the spatial infinity. It is this condition and the finiteness condition for $[H_\xi]$ that correspond virtually to the criterion for the applicability of the background subtraction method. Note that both conditions are now taking the form amenable to analysis at infinity. This represents a dramatic simplification in the Euclidean approach to black hole thermodynamics via the background subtraction method. In the prevailing strategy to check the validity of the first law of black hole thermodynamics via the background subtraction method, one is required to calculate out the black hole entropy through the explicit black hole solution, which is notoriously difficult to obtain especially for rotating black holes in the higher derivative gravity. But with our newly formulated criterion, one can circumvent this simply by examining the asymptotical behavior of the black hole solution at infinity, which is obviously much simpler. In addition, it is noteworthy that the resulting criterion does not require the isometric embedding, which not only offers a sharp explanation for the aforementioned applicability of the background subtraction method to some specific circumstances where the isometric embedding is impossible, but also opens up the possibility that the background subtraction method may share the same applicability as the covariant counterterm method in more generic scenarios. As we shall show in the next section, this is actually the case. 
    Now suppose that this criterion is satisfied, then we can further obtain the Gibbs free energy of the stationary black holes by going to the Euclidean sector as follows. First, by Eq. (\ref{Noethercurrent}), we have 
    \begin{equation}\label{Gibbs2}
        \beta\int_\infty \mathbf{Q}_\xi-S=-\beta\int_\Sigma\xi\cdot \mathbf{L}=\int_M d\tau\wedge \frac{\partial}{\partial\tau}\cdot\mathbf{L}_E=\int_M\mathbf{L}_E,
    \end{equation}
    where $\beta=\frac{2\pi}{\kappa}=\frac{1}{T}$ denotes the period along the Euclidean time $\tau=it$ and $\mathbf{L}_E=-i\mathbf{L}$. Similarly, we also have
    \begin{equation}\label{Gibbs1}
    \beta\int_\infty \mathbf{Q}^0_\xi=\int_{M^0}\mathbf{L}^0_E.
    \end{equation}
    Combining it with Eq. (\ref{Gibbs3}) as well as Eq. (\ref{Gibbs2}) and using the fact $X_\xi\cdot \mathbf{C}=0$, we end up with the expression of the Gibbs free energy as
    \begin{equation}
        \beta G\equiv\beta([H_\xi]-TS)=[I_E].
    \end{equation}
    Here the Euclidean action $I_E$ is given by
    \begin{equation}
        I_E=\int_M\mathbf{L}_E+\int_\infty \mathbf{B}_E
    \end{equation}
    with $\mathbf{B}_E=-i\mathbf{B}$, where the sign in front of the boundary integral comes from our previous convention for the orientations on $\Gamma$ and $\partial \Sigma_1$.
    
\section{Applicability to Einstein's gravity and its higher derivative corrections}
    Among others, below we shall demonstrate that the aforementioned criterion is satisfied by the four dimensional Einstein's gravity in both asymptotically flat and asymptotically AdS spacetimes as well as its higher derivative corrections, which not only provides the first evidence for the same applicability of the background subtraction method as the covariant counterterm method, but also paves the way for one to explore the higher derivative corrections to black hole thermodynamics via the background subtraction method, which is much simpler than the covariant counterterm method. 
    
For Einstein's general relativity, the Lagrangian form is given by 
    \begin{equation}\label{EH}
        \mathbf{L}=\frac{1}{16\pi}\bm{\epsilon}(R+\frac{6}{l^2}),
    \end{equation}
    where the cosmological constant $\Lambda=-\frac{3}{l^2}$. Whence we obtain $\psi_{abcd}=\frac{1}{32\pi}(g_{ac}g_{bd}-g_{ad}g_{bc})$ and $\Psi_{ab}=\frac{1}{32\pi}h_{ab}$. Accordingly, we have
    \begin{eqnarray}
        \mathbf{B}=\frac{K}{8\pi}\hat{\bm{\epsilon}},
        \quad \mathbf{C}=-\frac{1}{16\pi}\dbar{A}\cdot\hat{\bm{\epsilon}},\quad
          \mathbf{F}=-\frac{1}{2}T_{bc}\delta 
        h^{bc}\hat{\bm\epsilon},
        \quad q_\xi^a=T^a{}_c\xi^c,
    \end{eqnarray}
    where $T_{bc}= -\frac{1}{8\pi}(K_{bc}-Kh_{bc})$ is the corresponding Brown-York boundary energy-momentum tensor. 
    In addition, the corresponding rotating stationary black hole solution reads
    \begin{eqnarray}
        ds^2=-\frac{\Delta}{\rho^2}(dt-\frac{a}{\Xi}\sin^2\theta d\phi)^2+\frac{\rho^2}{\Delta}dr^2
        +\frac{\rho^2}{\Delta_\theta}d\theta^2+\frac{\Delta_\theta\sin^2\theta}{\rho^2}(adt-\frac{r^2+a^2}{\Xi}d\phi)^2
    \end{eqnarray}
    in the Boyer-Lindquist coordinates, 
    where 
    \begin{eqnarray}
       \Delta\equiv (r^2+a^2)(1+\frac{r^2}{l^2})-2mr+q^2,\quad \Delta_\theta\equiv 1-\frac{a^2}{l^2}\cos^2\theta,\quad 
        \rho^2\equiv r^2+a^2\cos^2\theta,\quad \Xi\equiv 1-\frac{a^2}{l^2}
    \end{eqnarray}
    with $q=0$\footnote{Actually all the following results in this section apply equally to the case with $q\neq 0$.}.
    In what follows, we shall denote $\Delta_0=\Delta|_{m=0}$ and choose $\Gamma$ to be the surface of $r=\bar{r}$. 
    
    To proceed, we make the following coordinate transformation 
\begin{equation}\label{shift1}
    t\rightarrow \frac{\sqrt{\Delta(\bar{r})}}{\sqrt{\Delta_0(\bar{r})}}t
\end{equation}
for the $m=0$ solution, which corresponds to the reference spacetime in the Boyer-Lindquist coordinates. Then the resulting metric for our reference spacetime reads
\begin{equation}
        ds^2=-\frac{\Delta_0}{\rho^2}(\frac{\sqrt{\Delta(\bar{r})}}{\sqrt{\Delta_0(\bar{r})}}dt-\frac{a}{\Xi}\sin^2\theta d\phi)^2+\frac{\rho^2}{\Delta_0}dr^2+\frac{\rho^2}{\Delta_\theta}d\theta^2+\frac{\Delta_\theta\sin^2\theta}{\rho^2}(a\frac{\sqrt{\Delta(\bar{r})}}{\sqrt{\Delta_0(\bar{r})}}dt-\frac{r^2+a^2}{\Xi}d\phi)^2,
    \end{equation}
whereby we like to choose the following basis
 \begin{eqnarray}
       &&e^0\equiv \sqrt{\frac{\Delta_0}{\rho^2}}(\frac{\sqrt{\Delta(\bar{r})}}{\sqrt{\Delta_0(\bar{r})}}dt-\frac{a}{\Xi}\sin^2\theta d\phi),\quad e^1\equiv \sqrt{\frac{\rho^2}{\Delta_0}}dr, \nonumber\\
       &&e^2\equiv \sqrt{\frac{\rho^2}{\Delta_\theta}}d\theta, \quad e^3\equiv\sqrt{\frac{\Delta_\theta\sin^2\theta}{\rho^2}}(a\frac{\sqrt{\Delta(\bar{r})}}{\sqrt{\Delta_0(\bar{r})}}dt-\frac{r^2+a^2}{\Xi}d\phi), 
    \end{eqnarray}
such that it is orthonormal with respect to our reference spacetime.

Next we shall present the relevant results for asymptotically flat and asymptotically AdS spacetimes, respectively.

\subsection{4D asymptotically flat spacetimes}

For the asymptotically flat case, the non-vanishing basis components of the induced metric and Brown-York tensor of the Kerr metric at $r=\bar{r}$ can be calculated as follows
\begin{eqnarray}
&& h_{00}=-1-\frac{2ma^2\sin^2\theta}{\bar{r}^3}+\mathcal{O}(\frac{1}{\bar{r}^4}), \quad h_{03}=\frac{2ma\sin\theta}{\bar{r}^2}+\mathcal{O}(\frac{1}{\bar{r}^3}), \nonumber\\
&& h_{22}=1, \quad h_{33}=1-\frac{2ma^2\sin^2\theta}{\bar{r}^3}+\mathcal{O}(\frac{1}{\bar{r}^4}),\nonumber\\
   &&T_{00}=-\frac{1}{4\pi\bar{r}}+\frac{m}{4\pi\bar{r}^2}+\mathcal{O}(\frac{1}{\bar{r}^3}),\quad T_{03}=\frac{a\sin\theta}{8\pi\bar{r}^2}+\frac{3ma\sin\theta}{8\pi\bar{r}^3}+\mathcal{O}(\frac{1}{\bar{r}^4}),\nonumber\\
   &&T_{22}=\frac{1}{8\pi\bar{r}}+\frac{m^2-a^2(1+\cos^2\theta)}{16\pi\bar{r}^3}+\mathcal{O}(\frac{1}{\bar{r}^4})=T_{33}
    \end{eqnarray}
with $m=0$ giving rise to the corresponding result for the reference  spacetime.  Although $\sqrt{|h|}=\frac{\rho\sqrt{\Delta}\sin\theta}{\Xi}$ is exactly the same for the Kerr and reference spacetime with the leading order given by $\sin\theta \bar{r}^2$, the induced metric of the Kerr black hole at $r=\bar{r}$
is not exactly the same as that of the reference spacetime for the background subtraction procedure. Instead, we have 
\begin{equation}\label{anotherbc}
    [h_{00}]=\mathcal{O}(\frac{1}{\bar{r}^3}),\quad [h_{03}]=\mathcal{O}(\frac{1}{\bar{r}^2}),\quad [h_{33}]=O(\frac{1}{\bar{r}^3}).
\end{equation}
In addition, for the first order perturbed Kerr solution, the non-vanishing basis components of the induced metrics can be calculated as 
\begin{eqnarray}
&& h'_{00}=-1+\frac{2\delta m}{\bar{r}}+\frac{4m\delta m}{\bar{r}^2}+\mathcal{O}(\frac{1}{\bar{r}^3}), \quad h'_{03}=\frac{2(ma+m\delta a)\sin\theta}{\bar{r}^2}+\mathcal{O}(\frac{1}{\bar{r}^3}),\nonumber\\
  && h'_{22}=1+\frac{2a\delta a\cos^2\theta}{\bar{r}^2}+\mathcal{O}(\frac{1}{\bar{r}^3}),\quad h'_{33}=1+\frac{2a\delta a}{\bar{r}^2}+\mathcal{O}(\frac{1}{\bar{r}^3}),\nonumber\\
   && h'^0_{00}=-1+\frac{2\delta m}{\bar{r}}+\frac{4m\delta m}{\bar{r}^2}+\mathcal{O}(\frac{1}{\bar{r}^3}), \quad h'^0_{03}=-\frac{2a\delta m\sin\theta}{\bar{r}^2}+\mathcal{O}(\frac{1}{\bar{r}^3}),\nonumber\\
  && h'^0_{22}=1+\frac{2a\delta a\cos^2\theta}{\bar{r}^2}+\mathcal{O}(\frac{1}{\bar{r}^3}),\quad h'^0_{33}=1+\frac{2a\delta a}{\bar{r}^2}+\mathcal{O}(\frac{1}{\bar{r}^3}).
\end{eqnarray}
It follows that the variation of the induced metric satisfies $\delta h_{\mu\nu}=\mathcal{O}(\frac{1}{\bar{r}})$ and $[\delta h_{\mu\nu}]=\mathcal{O}(\frac{1}{\bar{r}^2})$, whereby $\int_{\partial\Sigma_1}\xi\cdot[\mathbf{F}]=\mathcal{O}(\frac{1}{\bar{r}})$, ensuring that the condition (\ref{ourcriterion}) is fulfilled.

On the other hand,  the finiteness of $[H_\xi]$ through Eq. (\ref{simple}) is also satisfied according to the explicit expression for the following coordinate components of the Brown-York tensor
\begin{eqnarray}
T^t{}_t=\frac{1}{4\pi\bar{r}}-\frac{m}{4\pi\bar{r}^2}+\mathcal{O}(\frac{1}{\bar{r}^3}),\quad 
T^t{}_\phi=\frac{3ma\sin^2\theta}{8\pi\bar{r}^2}+\mathcal{O}(\frac{1}{\bar{r}^3})
\end{eqnarray}
with $m=0$ giving the corresponding result for the reference spacetime. Actually, whence one can further obtain the familiar ADM mass and angular momentum as follows
\begin{equation}
    M=m, \quad J=ma
\end{equation}
as it should be the case. 

Now let us check the potential effect on the applicability of the background subtraction method from the higher derivative terms.
To be more general, we assume that the form of the leading terms near the spatial infinity for the higher derivative corrected black hole solution is the same as that for the Kerr-Newman solution. Then an explicit calculation tells us that 
the basis components of extrinsic curvature and Riemann curvature have the following asymptotic behavior
\begin{equation}
K_{ab}=\mathcal{O}(\frac{1}{\bar{r}}),
\quad R_{abcd}=\mathcal{O}(\frac{1}{\bar{r}^3}).
\end{equation}
In addition, note that the covariant derivative always increases the order $\frac{1}{\bar{r}}$. Thus $\mathbf{F}$ from the higher derivative terms decays at least as $\mathcal{O}(\frac{1}{\bar{r}^2})$, which means that the higher derivative terms do not break down the first law of black hole thermodynamics.
On the other hand,  one can also show that the higher derivative terms do not contribute to the ADM mass and angular momentum through Eq. (\ref{simple}) as their contribution decays at least as $\mathcal{O}(\frac{1}{\bar{r}^2})$. Similarly, the higher derivative terms do not contribute to the boundary term $\mathbf{B}_E$ in the Euclidean action expression for the Gibbs free energy as their contribution also decays at least as $\mathcal{O}(\frac{1}{\bar{r}^2})$.

\subsection{4D asymptotically AdS spacetimes}

For the asymptotically AdS case, the whole analysis is the same, but much more involved with some subtleties and differences. First, before evaluating the metric perturbation towards the first order perturbed Kerr-AdS black hole solution, one is required to make the coordinate transformation $\phi\rightarrow \phi-\frac{\delta a}{l^2}t$ for the first order perturbed one such that the corresponding non-rotating observers at infinity share the same expression $\frac{\partial}{\partial t}-\frac{a}{l^2}\frac{\partial}{\partial \phi}$ as those in the unperturbed Kerr-AdS black hole. Second, in spite of the resulting $\int_{\partial\Sigma_1}\xi\cdot[\mathbf{F}]=\mathcal{O}(1)$, it turns out to be proportional to $\int_{-1}^1dx(1-3x^2)=0$ with $x=\cos\theta$, which guarantees the validity of the first law of black hole thermodynamics\footnote{Readers are kindly suggested to refer to the appendix for straightforward but more detailed calculations.}. 

On the other hand, according to 
\begin{eqnarray}\label{pure}
T^t{}_t=\frac{1}{4\pi l}+\frac{l^2-2a^2\cos^2\theta}{8\pi l \bar{r}^2 }-\frac{ml}{4\pi \bar{r}^3}+\mathcal{O}(\frac{1}{\bar{r}^4}),\quad 
T^t{}_\phi=\frac{3mal\sin^2\theta}{8\pi \Xi \bar{r}^3}+\mathcal{O}(\frac{1}{\bar{r}^4}),
\end{eqnarray}
one can obtain the finite ADM mass and angular momentum as usual
\begin{equation}
    M=\frac{m}{\Xi^2},\quad J=Ma.
\end{equation}

Next we shall examine the effect induced by the higher derivative terms out of Ricci scalar $R$, the traceless part of Ricci tensor $\mathcal{R}_{ab}$ and Weyl tensor $C_{abcd}$. First, it is noteworthy that the higher derivative terms may affect the AdS radius. To see this, let us consider the Einstein-Hilbert Lagrangian form supplemented by the higher derivative terms such as $\frac{\varepsilon}{16\pi}\bm{\epsilon}L_i$ with $i$ the total number of derivatives of metric and assume that it allows a pure AdS as its solution. Then we apply Eq. (\ref{variationradius}) to the perturbation of the pure AdS induced by $g_{ab}\rightarrow e^{2\lambda} g_{ab}$ with $\lambda$ a small constant parameter. As a result, one can obtain
\begin{equation}\label{adsradius}
    R+\frac{12}{l^2}+\varepsilon\frac{4-i}{2}L_i=0,
\end{equation}
where we have used $d\mathbf{\Theta}=0$. Note that $\mathcal{R}_{ab}=0$ and  $C_{abcd}=0$ for the pure AdS, so the correction to the AdS radius comes only from $R^n$ with $n>2$.

Except this potential correction, we shall assume that the leading form of the higher derivative corrected black hole solution is captured by the Kerr-Newman-AdS solution. Accordingly, we find that the basis components of the extrinsic curvature and Riemann curvature of the Kerr-Newman-AdS solution display the following asymptotic behavior 
\begin{equation}
    K_{ab}=\mathcal{O}(1),\quad R=\mathcal{O}(1),\quad \mathcal{R}_{ab}=\mathcal{O}(\frac{1}{\bar{r}^4}),\quad C_{abcd}=\mathcal{O}(\frac{1}{\bar{r}^3}).
\end{equation}
Whence the additional effect comes only from $R^n$, $R^mC^2$ as well as $R^m\tilde{C}^2$, where $C^2\equiv C_{abcd}C^{abcd}$ and $\tilde{C}^2\equiv C_{abcd}\tilde{C}^{abcd}$ with 
$\tilde{C}_{abcd}\equiv \frac{1}{2}\epsilon_{ab}{}^{ef}C_{efcd}$.

For $L_{2n}=R^n$, we have
\begin{equation}\label{admc1}
    \mathbf{B}=\frac{\varepsilon}{8\pi}nR^{n-1}K\hat{\bm{\epsilon}},\quad \mathbf{F}=\varepsilon[-\frac{1}{2}nR^{n-1}T_{bc}\delta h^{bc}+\frac{1}{8\pi}n(n-1)KR^{n-2}\delta R]\hat{\bm{\epsilon}}=-\frac{\varepsilon}{2}nR^{n-1}T_{bc}\delta h^{bc}\hat{\bm{\epsilon}},
    \quad q_\xi^a=\varepsilon nR^{n-1}T^a{}_c\xi^c,
\end{equation}
where we have used the fact $R=-\frac{12}{l^2}$ with $l$ understood as the corrected AdS radius through Eq. (\ref{adsradius}).
Thus the criterion for the applicability of the background subtraction method is obviously satisfied, although the ADM mass and angular momentum get corrected with both corrections equally proportional to the expression for the pure Einstein's gravity. Furthermore, note that
\begin{equation}
    K=\frac{3}{l}+\frac{a^2+l^2-5a^2\cos^2\theta}{2l\bar{r}^2}+\frac{l^4-2a^2l^2[1+3\cos^2\theta+a^4(1-6\cos^2\theta+21\cos^4\theta)]}{8l\bar{r}^4}+\frac{-ml(l^2\Delta_\theta+a^2)}{\bar{r}^5}+\mathcal{O}(\frac{1}{\bar{r}^6})
\end{equation}
with $m=0$ giving rise to the result for the pure AdS. This implies no boundary contribution to the Gibbs free energy via $[\mathbf{B}_E]$, which is the same as the case for the pure Einstein's gravity. 

For $L_{2(n+1)}=R^nC^2$, the potential contribution to $\mathbf{B}$ is given by 
\begin{equation}
\mathbf{B}\sim\frac{\varepsilon}{4\pi}\Phi_{ab}K^{ab}\hat{\bm{\epsilon}}=\mathcal{O}(\frac{1}{\bar{r}^2})
\end{equation}
with $\Phi_{ab}\equiv 2R^nC_{acbd}n^cn^d$. Thus there is also no boundary contribution to the Gibbs free energy. 
In addition, the potential contribution to $\mathbf{F}$ is given by 
\begin{equation}
\mathbf{F}\sim\frac{\varepsilon}{8\pi}(-\Phi_{ab}K^a{}_c\delta h^{bc}+2K_{bc}\delta \Phi^{bc})\hat{\bm{\epsilon}}=\mathcal{O}(1)\propto dx(1-3x^2),
\end{equation}
thus the integral of $\mathbf{F}$ over $\partial \Sigma_1$ vanishes, which guarantees the validity of the first law of black hole thermodynamics. However, the ADM mass and angular momentum will get corrected, because the potential contribution to the Brown-York boundary energy momentum tensor is given by 
\begin{equation}
    \mathcal{T}^a{}_c\sim-\frac{\varepsilon}{8\pi}(3\Phi^{ab}K_{cb}-K^{ab}\Phi_{cb})=\mathcal{O}(\frac{1}{\bar{r}^3}).
\end{equation}
To be more specific, we obtain
\begin{equation}\label{admc2}
    \mathcal{T}^t{}_t=-\frac{\varepsilon R^nm}{\pi l\bar{r}^3} +\mathcal{O}(\frac{1}{\bar{r}^4}),\quad \mathcal{T}^t{}_\phi=\frac{\varepsilon R^n3ma\sin^2\theta}{2\pi l\Xi\bar{r}^3}+\mathcal{O}(\frac{1}{\bar{r}^4}).
\end{equation}
By comparing the above equation with Eq. (\ref{pure}), we find 
that the corrections to the ADM mass and angular momentum are also equally proportional to the expression for the pure Einstein's gravity with the proportional coefficient given by $\frac{\varepsilon 4R^n}{l^2}$.

Last, for $L_{2(n+1)}=R^n\tilde{C}^2$, we have $\tilde{\Phi}_{ab}\equiv R^n\tilde{C}_{acbd}n^cn^d=\mathcal{O}(\frac{1}{\bar{r}^4})$. This implies that this higher derivative term does not affect the applicability of the background subtraction method at all.

Here is the punchline for the AdS case. Although the higher derivative terms may have a bearing on the expression for the ADM mass and angular momentum in a miraculously simple manner, they do not invalidate the background subtraction method, whereby the contribution to the Gibbs free energy comes solely from the bulk Euclidean action.



\section{Conclusion}
As the criterion for the applicability of the background subtraction method, not only the finiteness condition of the resulting Hamiltonian, but also the condition for the validity of the first law of black hole thermodynamics boils down to the form amenable to examination at infinity. Among others, one very advantage of this finding over the prevalent strategy to check the validity of the first law is that one is not required any more to calculate the black hole entropy, which depends sensitively on the full metric deep into the bulk. Instead, it is sufficient for one to analyze the asymptotical behavior of 
the metric near the infinity, which turns out to be much easier. 
With this criterion, we further establish that the background subtraction method is applicable not only to Einstein's gravity but also to its higher derivative corrections for black hole thermodynamics in both asymptotically flat and AdS spacetimes. As a bonus, we also obtain the universal expression of the Gibbs free energy in terms of the Euclidean on-shell action, which has been derived before via the strategy special to Einstein's gravity but assumed to be also valid beyond Einstein's gravity without derivation \cite{RS}.

Although we only show that the background subtraction method is applicable to 4D gravity, we believe that it is also applicable to other dimensions. In addition, we are convinced that the whole framework developed here using the covariant phase space formalism can be applied to address the applicability of the background subtraction method to more generic scenarios we probably encounter. In this sense, the background subtraction method is arguably as applicable as the covariant counterterm method, correcting the everlasting bias that it is rather restrictive compared to the latter.
Taking into account that the background subtraction method is much simpler than the covariant counterterm method, our result establishes a solid theoretical foundation for one to apply the former to greatly facilitate the otherwise much more involved calculations in the Euclidean approach to black hole thermodynamics by the latter \cite{Ma,Hu,Ma2,Cremonini,Belgium}.

\begin{acknowledgments}
This work is partially supported by the National Key Research and Development Program of China with Grant No. 2021YFC2203001 as well as the National Natural Science Foundation of China with Grant Nos. 12075026 and 12361141825. 
In addition, this work was done in part during the workshop ``Holographic Duality and Models of Quantum Computation" held at Tsinghua Southeast Asia Center on Bali, Indonesia (2024). 
HZ is grateful to Pengju Hu, Hong Lu, Liang Ma, Yi Pang, and Yong Xiao for sharing their wisdom on higher derivative corrections. He would also like to thank Yu Tian, Jieqiang Wu, and Amos Yarom for their stimulating discussions on this work.

\end{acknowledgments}

\newpage

\onecolumngrid
\appendix

\section{The validity of the first law of black hole thermodynamics for Kerr-AdS in Einstein's gravity}

For the Kerr-AdS metric, we have 
\begin{eqnarray}
&& h_{00}=-1-\frac{2ma^2l^2\sin^2\theta}{\bar{r}^5}+\mathcal{O}(\frac{1}{\bar{r}^6}), \quad h_{03}=\frac{mal\sin\theta}{\sqrt{\Delta_{\theta}}\bar{r}^3}+\mathcal{O}(\frac{1}{\bar{r}^4}), \nonumber\\
&&h_{22}=1, \quad h_{33}=1-\frac{2ma^2l^2\sin^2\theta}{\bar{r}^5}+\mathcal{O}(\frac{1}{\bar{r}^6}),\nonumber\\
    &&T_{00}=-\frac{1}{4\pi l}-\frac{a^2+l^2-3a^2\cos^2\theta}{8\pi l\bar{r}^2}+\frac{ml}{4\pi\bar{r}^3}+\mathcal{O}(\frac{1}{\bar{r}^4}),\quad T_{03}=\frac{a\sin\theta\sqrt{\Delta_{\theta}}}{8\pi\bar{r}^2}+\frac{ma\sin\theta}{4\pi\sqrt{\Delta_{\theta}}\bar{r}^3}+\mathcal{O}(\frac{1}{\bar{r}^4}),\nonumber\\
    && T_{22}=\frac{1}{4\pi l}-\frac{a^2\cos^2\theta}{8\pi l\bar{r}^2}+\frac{ml}{8\pi \bar{r}^3}+\mathcal{O}(\frac{1}{\bar{r}^4})=T_{33}
\end{eqnarray}
with $m=0$ giving rise to the corresponding result for the reference spacetime, whereby 
\begin{equation}\label{bcanother}
    [h_{00}]=\mathcal{O}(\frac{1}{\bar{r}^5}),
\quad [h_{03}]=\mathcal{O}(\frac{1}{\bar{r}^3}),\quad [h_{33}]=\mathcal{O}(\frac{1}{\bar{r}^5}).
\end{equation}
In addition, for the first order perturbed Kerr-AdS solution, we are required to first make the following coordinate transformation 
\begin{equation}
t=t', \quad \phi=\phi'+\frac{\delta a}{l^2}t',\quad r=r',\quad \theta=\theta',
\end{equation}
such that the non-rotating observers at infinity $\frac{\partial}{\partial t'}+\Omega'_\infty\frac{\partial}{\partial \phi'}=\frac{\partial}{\partial t}+\Omega_\infty\frac{\partial}{\partial \phi}$ with $\Omega_\infty=-\frac{a}{l^2}$ share the same expression in the unprimed coordinates as those in the unperturbed black hole background. Accordingly, we have 
\begin{eqnarray}
 h'_{00}&=&-1-\frac{2a\delta a \sin^2\theta}{l^2\Xi}+\frac{2a^3\delta a\sin^2\theta(1+\cos^2\theta)}{l^2\Xi \bar{r}^2}+\frac{2l^2\delta m}{\bar{r}^3}+\mathcal{O}(\frac{1}{\bar{r}^4}), \nonumber\\
 h'_{03}&=&-\frac{2a^2\delta a\cos^2\theta \sin\theta }{l^3\Xi \sqrt{\Delta_\theta}}+\frac{a^2\delta a \sin\theta(-2l^2-l^2\Xi \cos^2\theta+2a^2\cos^4\theta)}{l^3\Xi \sqrt{\Delta_\theta}\bar{r}^2}\nonumber\\
&&+\frac{al^2\Xi m\sin\theta+m\delta a \sin\theta(a^2+2l^2-a^2\cos2\theta)}{l\Xi \sqrt{\Delta_\theta}\bar{r}^3}+\mathcal{O}(\frac{1}{\bar{r}^4}),\nonumber\\
 h'_{22}&=&1+\frac{2a\delta a \cos^2\theta}{l^2\Delta_\theta}+\frac{2a\delta a\cos^2\theta}{\bar{r}^2}+\mathcal{O}(\frac{1}{\bar{r}^4}),\nonumber\\
  h'_{33}&=&1+\frac{2a\delta a (l^2-a^2\cos^4\theta)}{l^4\Xi \Delta_\theta}+\frac{2a\delta a(l^2-a^2\cos^4\theta)}{l^2\Xi \bar{r}^2}+\frac{a^3m\delta a\sin^22\theta}{l^2\Xi\Delta_\theta\bar{r}^3}+\mathcal{O}(\frac{1}{\bar{r}^4}),\nonumber\\
  h'^0_{00}&=&-1-\frac{ 2a \delta a\sin^2\theta}{l^2\Xi}+\frac{2a^3\delta a\sin^2\theta(1+\cos^2\theta)}{l^2\Xi \bar{r}^2}+\frac{2l^2\Xi\delta m+ 2am\delta a\sin^2\theta}{\Xi \bar{r}^3}+\mathcal{O}(\frac{1}{\bar{r}^4}), \nonumber\\
   h'^0_{03}&=&-\frac{2a^2\delta a\cos^2\theta \sin\theta }{l^3\Xi \sqrt{\Delta_\theta}}+\frac{a^2\delta a \sin\theta(-2l^2-l^2\Xi \cos^2\theta+2a^2\cos^4\theta)}{l^3\Xi \sqrt{\Delta_\theta}\bar{r}^2}\nonumber\\
   &&+\frac{-al^2\Xi\delta m\sin\theta+m\delta a \sin\theta(l^2-a^2\cos2\theta)}{l\Xi \sqrt{\Delta_\theta}\bar{r}^3}+\mathcal{O}(\frac{1}{\bar{r}^4}),\nonumber\\
   h'^0_{22}&=&1+\frac{2a\delta a\cos^2\theta}{l^2\Delta_\theta}+\frac{2a\delta a\cos^2\theta}{\bar{r}^2}+\mathcal{O}(\frac{1}{\bar{r}^4}),\nonumber\\
  h'^0_{33}&=&1+\frac{2a\delta a (l^2-a^2\cos^4\theta)}{l^4\Xi \Delta_\theta}+\frac{2a\delta a(l^2-a^2\cos^4\theta)}{l^2\Xi \bar{r}^2}+\frac{2am\delta a\sin^2\theta}{\Xi \bar{r}^3}+\mathcal{O}(\frac{1}{\bar{r}^4})
\end{eqnarray}
for the first order perturbed Kerr-AdS solution in the unprimed coordinates. By $\delta h^{ab}=-h^{ac}h^{bd}\delta h_{cd}$, we further have
\begin{eqnarray}
 \delta{h'^{00}}&=&\frac{2a\delta a\sin^2\theta}{l^2\Xi}-\frac{2a^3\delta a\sin^2\theta(1+\cos^2\theta)}{l^2\Xi\bar{r}^2}-\frac{ma^3\delta a\sin^22\theta}{l^2\Xi \Delta_\theta\bar{r}^3}-\frac{2l^2\delta m}{\bar{r}^3}+\mathcal{O}(\frac{1}{\bar{r}^4}), \nonumber\\
\delta{h'^{03}}&=&-\frac{2a^2\delta a\sin\theta\cos^2\theta}{l^3\Xi\sqrt{\Delta_\theta}}+\frac{a^2\delta a\sin\theta(-2l^2-l^2\Xi \cos^2\theta+2a^2\cos^4\theta)}{l^3\Xi\sqrt{\Delta_\theta}\bar{r}^2}\nonumber\\
&&+\frac{2m\delta a\sin\theta(l^4\Xi-a^2l^2\cos^2\theta+a^4\cos^4\theta)}{l^3\Xi\Delta_\theta\sqrt{\Delta_\theta} \bar{r}^3}+\mathcal{O}(\frac{1}{\bar{r}^4}), \nonumber\\
\delta{h'^{22}}&=&-\frac{2a\delta a cos^2\theta}{l^2\Delta_\theta}-\frac{2a\delta a\cos^2\theta }{\bar{r}^2}+\mathcal{O}(\frac{1}{\bar{r}^4}), \nonumber\\
\delta{h'^{33}}&=&-\frac{2a\delta a(l^2-a^2\cos^4\theta)}{l^4\Xi \Delta_\theta}-\frac{2a\delta a(l^2-a^2\cos^4\theta)}{l^2\Xi\bar{r}^2}+\mathcal{O}(\frac{1}{\bar{r}^4}),\nonumber\\
 \delta{h'^{0}}^{00}&=&\frac{2a\delta a\sin^2\theta}{l^2\Xi}-\frac{2a^3\delta a\sin^2\theta(1+\cos^2\theta)}{l^2\Xi\bar{r}^2}-\frac{2ma\delta a\sin^2\theta}{\Xi \bar{r}^3}-\frac{2l^2\delta m}{\bar{r}^3}+\mathcal{O}(\frac{1}{\bar{r}^4}), \nonumber\\
\delta{h'^{0}}^{03}&=&-\frac{2a^2\delta a\sin\theta\cos^2\theta}{l^3\Xi\sqrt{\Delta_\theta}}+\frac{a^2\delta a\sin\theta(-2l^2-l^2\Xi \cos^2\theta+2a^2\cos^4\theta)}{l^3\Xi\sqrt{\Delta_\theta}\bar{r}^2}\nonumber\\
&&+\frac{-al^2\Xi\delta m\sin\theta+m\delta a \sin\theta(l^2-a^2\cos2\theta)}{l\Xi \sqrt{\Delta_\theta}\bar{r}^3}+\mathcal{O}(\frac{1}{\bar{r}^4}), \nonumber\\
\delta{h'^{0}}^{22}&=&-\frac{2a\delta a \cos^2\theta}{l^2\Delta_\theta}-\frac{2a\delta a\cos^2\theta }{\bar{r}^2}+\mathcal{O}(\frac{1}{\bar{r}^4}), \nonumber\\
\delta{h'^{0}}^{33}&=&-\frac{2a\delta a(l^2-a^2\cos^4\theta)}{l^4\Xi \Delta_\theta}-\frac{2a\delta a(l^2-a^2\cos^4\theta)}{l^2\Xi\bar{r}^2}+\frac{2ma\delta a\sin^2\theta}{\Xi\bar{r}^3}+\mathcal{O}(\frac{1}{\bar{r}^4}).
\end{eqnarray}
Then a straightforward calculation shows that $\int_{\partial\Sigma_1}\xi\cdot[\mathbf{F}]$ is proportional to $\int_{-1}^1dx(1-3x^2)=0$ with $x=\cos\theta$, which ensures the validity of the first law of black hole thermodynamics.

\end{document}